\documentclass[twocolumn,showpacs,amsmath,amssymb,prl,superscriptaddress,floatfix]{revtex4}

\usepackage{graphicx}% Include figure files
\usepackage{dcolumn}% Align table columns on decimal point
\usepackage{bm}% bold math
\newcommand{\bg}[1]{\mbox{\boldmath$#1$}} %% bold greek
\usepackage{xcolor}
\definecolor{red}{rgb}{0.7,0,0}%darkred MIT
\definecolor{green}{rgb}{0.,0.35,0.}%darkgreen
\definecolor{blue}{rgb}{0.2,0.2,0.7} %beamer@blendedblue
\definecolor{black}{rgb}{0.15,0.15,.15}%not too black

\begin{document}

\title{Mesoscopic  phases of dipolar ensembles with polar molecules and Rydberg atoms }
\pacs{}

\author{G. Pupillo}
\affiliation{Institute for
Theoretical Physics, University of Innsbruck, and Institute for
Quantum Optics and Quantum Information of the Austrian Academy of
Sciences, Innsbruck, Austria}
\author{A. Micheli}
\affiliation{Institute for
Theoretical Physics, University of Innsbruck, and Institute for
Quantum Optics and Quantum Information of the Austrian Academy of
Sciences, Innsbruck, Austria}
\author{M. Boninsegni}
\affiliation{Department of Physics, University of Alberta, Edmonton, Alberta, Canada T6G 2J1}
\affiliation{Institute for
Theoretical Physics, University of Innsbruck, and Institute for
Quantum Optics and Quantum Information of the Austrian Academy of
Sciences, Innsbruck, Austria}
\author{I. Lesanovsky}
\affiliation{School of Physics and Astronomy, University of Nottingham, Nottingham, UK}
\author{P. Zoller}
\affiliation{Institute for Theoretical Physics, University of
Innsbruck, and Institute for Quantum Optics and Quantum Information
of the Austrian Academy of Sciences, Innsbruck, Austria}
\begin{abstract}
We discuss the realization of mesoscopic phases of dipolar gases relevant to current experiments with cold polar molecules and Rydberg atoms confined to two dimensions. We predict the existence of superfluid clusters, mesoscopic supersolids, and crystals for a small number of trapped particles, with no counterpart in the homogeneous situation. For certain strengths of the dipole-dipole interactions, the stabilization of purely {\it non-classical crystals} by quantum fluctuations is possible. We propose a magnification scheme to detect the spatial structure of these crystalline phases.
\end{abstract} \maketitle

One of the most remarkable achievements in recent AMO experiments is
the preparation of cold ensembles of polar molecules in the electronic
and rovibrational ground state~\cite{PolMolBook,PolMolExp}. The
distinctive feature of polar molecules is their comparatively large
electric dipole moments (of up to a few Debye) implying strong dipolar
interactions. These interactions can be manipulated via external DC
and AC microwave fields by coupling to excited rotational levels of
the molecule. In combination with reduced trapping geometries this
opens the door to a study of strongly correlated quantum phases with
designed long range interactions, e.g., for Bose systems the
superfluid-crystal quantum phase transition in 2D as a function of an
induced dipole moment~\cite{Buechler07}.  While ground state Alkali atoms exhibit much weaker magnetic dipole moments~\cite{ChromiumExp}, huge electric dipole moments $d \sim n^2$ are present in their high-lying Rydberg states, where $n$ is the principal quantum number~\cite{GallagherBook,RydbergExp}. Thus, below we propose that strong correlations can be observed in an atomic gas of ground state atoms
by weakly admixing with laser light these Rydberg states. In contrast to polar
molecules, these dressed atomic gases exhibit decoherence and heating
mechanisms from spontaneous emission  ($\Gamma \sim 1/n^3$) and
inelastic collisions. However, for strongly correlated phases,
e.g. in the crystalline phase,  the  character of the many body
wavefunction can strongly suppress the probability for close-encounter, possibly harmful, inelastic collisions, and thus stabilize these phases.

\begin{figure}
\includegraphics[width=0.8\columnwidth]{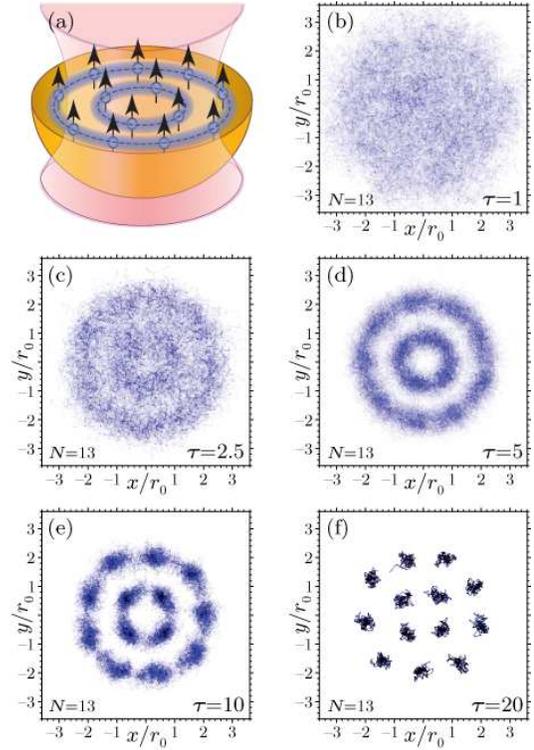}%[angle=0,width=7.5cm]{./pot_curve3.eps}
\caption{(color online) (a) Experimental setup (scheme): polar molecules or Rydberg-dressed atoms are confined to 2D by a strong confining laser beam, with dipoles polarized perpendicular to the plane. In-plane harmonic confinement is provided by, e.g., the beam waist. (b-e) Monte Carlo snapshots of the density of particles in all mesoscopic phases for $N=13$ dipoles, as a function of the effective mass $\tau$. (b) superfluid; (c) supersolid; (d-e) ring-like crystals; (f) classical crystal.}\label{fig:fig1}
\end{figure}

In light of such new experimental developments, we study in this letter  {\it mesoscopic quantum phases} and crossover between these phases in small clusters of dipolar particles confined to 2D, in the presence of an in-plane  harmonic confinement [see Fig.~\ref{fig:fig1}(a)]. By means of Quantum Monte Carlo simulations based on the continuous-space Worm Algorithm \cite{worm}, we show the existence of quantum phases which have no counterpart in the homogeneous situation. These include superfluid clusters, mesoscopic supersolids, ring-shaped crystals, and classical crystals with unusual rotational symmetries. We find that for certain particle numbers and intermediate interaction strength quantum fluctuation can stabilize purely non-classical crystals. Finally, we propose schemes to detect these crystalline phases.

We consider a setup where $N$ dipolar bosonic particles are confined to a 2D plane by applying a strong transverse trapping field~\cite{Buechler07,Micheli07}, e.g a 1D optical lattice, and are aligned perpendicular to the plane, with a DC induced dipole moment $d\equiv \sqrt{D} $. We assume an additional in-plane parabolic trap with frequency $\omega$, as realized by a magnetic dipole trap, or a single site of a large spacing optical lattice. For pure dipolar interactions the 2D Hamiltonian for $N$ particles with mass $m$ is
\begin{eqnarray}\label{eq:eqHam}
  H_{\rm 2D} = \sum_{i=1}^N \left[\frac{{\bf p}_i^2}{2m} + m\omega^2\frac{{\bf
        r}_i^2}{2}\right] + \sum_{i>j}\frac{D}{|{\bf r}_i-{\bf r}_j|^3},
\end{eqnarray}
which is the sum of kinetic energy, parabolic confinement, and dipole-dipole interactions. For polar molecules, the validity of Hamiltonian~\eqref{eq:eqHam} has been discussed in Refs.~\cite{Buechler07,Micheli07}; we will return to a discussion of its validity for Rydberg atoms and of relevant experimental parameters towards the end of the paper. Defining length and energy scales
$r_0=(D/m\omega^2)^{1/5}$ and $\epsilon_0=m\omega^2r_0^2=D/r_0^3=(m^3\omega^6D^2)^{1/5}$, respectively, Eq.~\eqref{eq:eqHam} can be recast in dimensionless form as
\begin{eqnarray}\label{eq:eqHamRes}
\frac{H}{\epsilon_0}=\sum_{i=1}^{N} \left[-\frac{1}{2 \tau^2}\frac{\partial^2}{\partial {\bg \rho}_i^2}+\frac{1}{2}{\bg \rho}_i^2\right]
    + \sum_{i>j}\frac{1}{|{\bg \rho}_i-{\bg \rho}_j|^3},
\end{eqnarray}
where $ \tau \equiv \epsilon_0 / \hbar \omega=\left(r_0/ \ell\right)^2 = \left(m D/ \hbar^2 \ell\right)^{2/5}$ characterizes the strength of the dipole-dipole interactions in the trap with $\ell = \sqrt{\hbar/m\omega}$ the harmonic oscillator length. Equation~\eqref{eq:eqHamRes} shows that $\tau$ plays the role of an {\em effective mass} representing a control parameter, which can be increased by increasing the strength of dipole-dipole interactions, or by compressing the trap. 

All the results shown here correspond to a low enough temperature $T$, that they can be regarded essentially ground state estimates. Quantitatively, this means working at a temperature (much) smaller than the characteristic energy scale $\epsilon_0$.
For small $\tau \lesssim 1$, we expect the kinetic energy to dominate, and the cluster to be in a weakly interacting superfluid phase. In the limit of strong interactions (or equivalently, large particle mass) $\tau \gg 1$, kinetic energy becomes negligible, and the system ground state resembles the classical lowest-energy crystalline configuration obtained by minimizing the last two terms of Eq.~\eqref{eq:eqHamRes}.

\begin{figure}
\includegraphics[width=0.9\columnwidth]{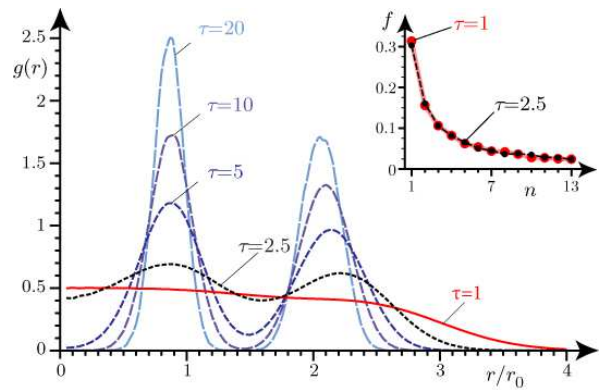}%[angle=0,width=7.5cm]{./pot_curve3.eps}
\caption{(color online) Radial density profiles $g(r)$ for the cases of Fig.~\ref{fig:fig1}(b-f). Inset: statistics of computed particle exchanges $f$ as a function of the number of particles participating  to the exchange $n$, for the cases of finite superfluid fraction $\tau=1$ and 2.5. The supersolid phase with $\tau=2.5$ has finite density modulation (figure) and the same $f$-distribution of the superfluid with $\tau=1$ (inset).}\label{fig:fig2}
\end{figure}
We obtain an estimate of the critical $\tau_{\rm c}$ for the crossover from the superfluid to the crystal by noting that for a homogeneous system the superfluid-crystal transition occurs at $r_{\rm QM}= D m/\hbar^2 a= 18 \pm 4$~\cite{Buechler07,Astrakharchik07}, where $r_{\rm QM}$ represents the ratio of the dipolar interactions $D/a^3$ to the kinetic energy $\hbar^2/ma$ with $a$ the mean interparticle distance. By rewriting $\tau_{\rm c} = (r_{\rm QM} a / \ell)^{2/5}$, and approximating $a \sim \ell$, we obtain the prediction $\tau_{\rm c} \simeq 3$, which is essentially $N$-{\it independent}. Figure~\ref{fig:fig1}(b-f) shows our results for $\tau$ ranging between 1 and 20 and $N=13$ particles, which we chose since we found that it displays all general features of mesoscopic clusters with $N \lesssim 30$. Panels (b-f) are snapshots of the particle density obtained by tracing over the worldline imaginary-time evolution in the Monte-Carlo simulations.
Consistently with our model estimate, panel (b) shows that for $\tau =1$ the system is in a superfluid phase, characterized by a flat, featureless, density profile, and superfluid fraction close to 100\%. The corresponding radial density profile $g(r)$ is shown in Fig.~\ref{fig:fig2}. For $\tau \gtrsim 3$ [panels (d-f)] the overlap between the various particle probability clouds rapidly drops to zero. 
In particular, for $\tau=20$ we find essentially no overlap, as particles are arranged so as to mimic the classical crystalline configuration for $N=13$, with 4 particles at the center, and 9 outside. In this case, Fig.~\ref{fig:fig2} shows two large peaks at the radial position $r$ corresponding to the classical equilibrium configuration for the crystal, and no density in between the peaks~\cite{note}.\\

The competition between interactions and confinement for intermediate values of $\tau$ allows for the existence of mesoscopic phases with no counterpart in the homogeneous situation. We find that for dipolar particles these are: {\it i)} a supersolid phase with superfluid fraction one, and finite density modulation for $\tau \lesssim \tau_{\rm c}$, and {\it ii)} ring-shaped crystals with no detectable superfluid fraction for $\tau \gtrsim \tau_{\rm c}$. We have checked that these mesoscopic phases, and in particular the supersolid, occur for all particle numbers $N<30$.

The emergence of supersolid-type behavior for $\tau \lesssim \tau_{\rm c}$ for $N=13$ is signaled by a density modulation on top of the superfluid background in Fig.~\ref{fig:fig1}(c). This is shown quantitatively in Fig.~\ref{fig:fig2}, where the radial density profile is clearly more structured compared to the case $\tau=1$. We find that the superfluid properties are unaltered by the increased strength of interactions. In fact, the superfluid fraction for $\tau=2.5$ approaches unity in the $T\to 0$ limit, and, as shown in the inset of Fig.~\ref{fig:fig2}, no difference is observed in the statistics of exchange cycles for $\tau=1$ and 2.5, computed at corresponding low temperatures (i.e., same fractions of $\epsilon_0$).

\begin{table}
\begin{tabular}{c|c|c|c|c|c}
  $N$ & $s$ & $\mu_{N,s}$ & $\nu_{N,s}$ & $\tau_{\star}$ & rings \\
  \hline
  \hline
  5 & 0 & 1.15927 & 1.45437 & 0      & (0,5) \\
  5 & 1 & 1.18902 & 1.42189 & 1.0917 & (1,4) \\%
%  \hline
%  10 & 0 & 2.10469 & 1.83091 & 0      & (0,3,(2,2,2,1))) \\
%  10 & 1 & 2.10548 & 1.87850 & -47.8471 & (0,2,2,4,2) \\
  \hline
  12 & 0 & 2.41909 & 2.02725 & 0 & (3,3,6) \\
  12 & 1 & 2.42054 & 2.00556 & 14.957 & (4,4,4)\\
  \hline
  19 & 0 & 3.41759 & 2.42749 & 0 & (1,6,6,6) \\
  19 & 1 &3.42138 & 2.39173 & 9.4275 & (1,7,11) \\
\end{tabular}
\caption{Parameters for possible non-classical/classical crystal transitions. Columns one to six indicate the number of particles $N$, the label $s$ for the classical (0) and non-classical (1) crystal configuration, the dimensionless quantities $\mu_{N,s}$ and $\nu_{N,s}$ of Eq.~\eqref{eq:eqSemEnergy}, the transition parameter $\tau_\star$ (see text), and the ring-configuration, respectively.}\label{table:table1}
\end{table}
For $\tau \gtrsim \tau_{\rm c}$ we observe crystallization of the atomic/molecular cloud, consistently with the estimates above. This is signaled by a drop of the superfluid fraction above $\tau = 3$.
We find that crystallization in these finite, inhomogeneous, systems proceeds through intermediate crystal-like phases, which are characterized by the arrangement of particles in concentric rings with a {\it fixed} number of particles per ring~\cite{note:Kivelson}. For small enough temperatures, these rings are free to rotate independently of each other. Snapshots of these ring-shaped phases are shown in Fig.~\ref{fig:fig1}(d-e) for $\tau=5$ and 10, respectively. These two cases differ only for the amount of localization of the particle wave-function in the rings. They both display no superfluid properties~\cite{note}.

The crystal-like ring-configurations above characterize the generic crossover between superfluid and classical crystal in mesoscopic clusters with $N \lesssim 30$. In the following we focus on a few special $N$, for which the crossover is characterized by the presence of a mesoscopic analog of a {\it first-order} transition between a {\it non-classical crystalline ground-state} and the classical one. These crystals differ with respect to rotational symmetry. To our knowledge, this is the first demonstration of the existence of these phases in mesoscopic clusters.

\begin{figure}
\includegraphics[width=0.9\columnwidth]{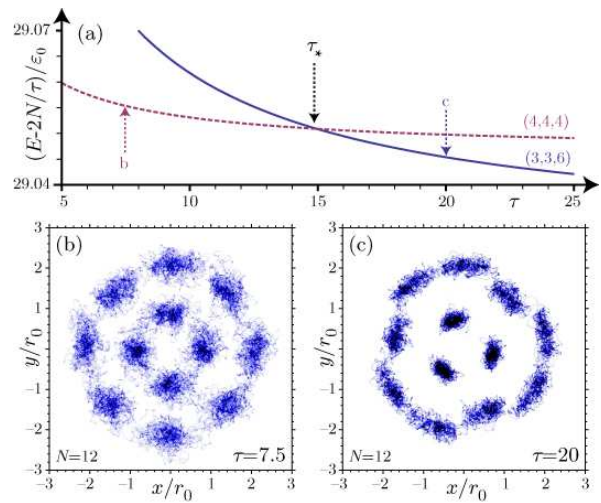}%{Figs/Fig1_AM_GP.eps}%[angle=0,width=7.5cm]{./pot_curve3.eps}
\caption{(color online) Non-classical vs classical crystals for $N=12$. (a) Energy as a function of $\tau$ as computed with Eq.~\eqref{eq:eqSemEnergy} for the configurations (4,4,4) (red dashed line) and (3,3,6) (blue continuous line), corresponding to the non-classical and the classical crystals, respectively. (b) and (c): Monte Carlo snapshots of the groundstate configuration for $\tau=7.5 < \tau_{\star}$ and $\tau=20 > \tau_\star$, respectively.}\label{fig:fig3}
\end{figure}

\begin{figure}[b]
\includegraphics[width=0.9\columnwidth]{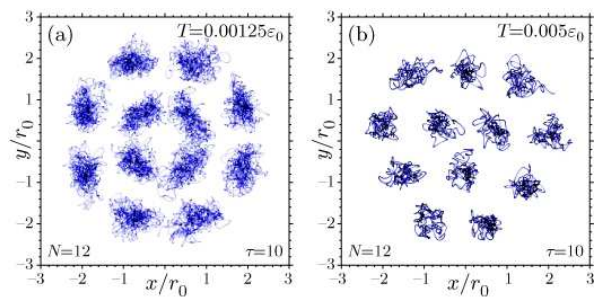}%[angle=0,width=7.5cm]{./pot_curve3.eps}
\caption{Finite temperature transition from the non-classical to the classical crystal for $\tau=10$ and $N=12$. (a) and (b) are Monte Carlo snapshots for $T=0.00125 \epsilon_0$ and $0.005 \epsilon_0$, respectively.}\label{fig:fig4}
\end{figure}
In general, for a given $N$ there can be several low-energy classical equilibrium configurations, which differ from the ground-state in the number of particles belonging to each ring~\cite{Belousov00}. These low-energy configurations are found, e.g., using a nonlinear unconstrained minimization of the potential and interaction energy terms in Eq.~\eqref{eq:eqHamRes} based on random initial configurations ${\bg \rho}_i$. For each configuration, low-energy excitations can be obtained semi-classically by computing the phonons, corresponding to small vibrations around the mean particle positions. For a particular configuration $s$ ($s=0$ and 1 are the classical and non-classical configurations, respectively) and $N$ particles, the mean-energy per particle $E_{N,s}$ reads
\begin{eqnarray}\label{eq:eqSemEnergy}
\frac{E_{N,s}}{N} = \epsilon_0 \mu_{N,s} + \hbar \omega \nu_{N,s}.
\end{eqnarray}
Here, $\mu_{N,s}$ is the dimensionless classical potential energy per
particle, while $\nu_{N,s}=1/N \sum_{k=1}^{2N} \omega_k/2\omega$ gives the ratio of mean quantum fluctuations to trap energy spacing. While the classical ground-state configuration has always the lowest energy for $\tau \gg 1$, it is in principle possible that for certain $N$ and intermediate $\tau$, quantum fluctuations stabilize as a ground-state a {\it crystalline} configuration  (e.g., no superfluidity) with a rotational symmetry which is different from the classical ground-state. 
A {\it semiclassical estimate} for this transition is obtained by the relation $\tau_{\star} = (\nu_1-\nu_0)/(\mu_0-\mu_1)$. Consistency with the requirement of a crystalline configuration implies $\tau_{\rm c}\lesssim \tau <  \tau_{\star}$.
Table 1 shows numerical values for a few favorable $N$, where this process could be visible.
Figure~\ref{fig:fig2} shows our results for $N=12$. Fully consistent with our estimates, we find that in the range $ 3 \lesssim \tau \lesssim 15$ the ground-state is the ring structure (4,4,4), [panel (b)], while the classical crystal configuration (3,3,6) sets in for $\tau > 15$, [panel (d)]. These two configurations have rotational symmetry $\mathcal{C}_4$ and $\mathcal{C}_3$, respectively. Actually, quantum mechanical fluctuations modify the ring structure (4,4,4) into two concentric rings with 4 and 8 particles, respectively. The radial density profiles for some $\tau$ are shown in panel (d). Most interestingly, we find that in the intermediate range of $\tau$ where the non-classical crystal is the ground-state, the classical ground-state can survive as a low-energy {\em metastable} configuration. It is then possible to induce a non-classical/classical-crystal transition by raising the temperature for fixed $\tau$. This is shown for $N=12$ and $\tau = 10$ in Fig.~\ref{fig:fig3}. Further melting into a featureless normal fluid occurs at even higher temperatures, which we find consistent with the classical melting temperature for the homogeneous crystal configuration~\cite{Kalia81} (not shown). Again, consistent with our semi-classical model, we found that the non-classical configuration (1,4) is not realized for $N=5$, since $\tau_\star \simeq 1 < \tau_{\rm c}$.\\

The mesoscopic phases above can be realized with polar molecules of current experimental interest. For example, for a moderate in-plane confinement $\omega / 2 \pi=$ 1kHz, $r_0=0.4, 0.8$ and 1 $\mu$m for RbCs ($d=1.25$Debye), LiCs ($d=5.5$Debye), and SrO molecules ($d=8.9$Debye),  respectively, and $\tau \simeq 4, 9$ and 11. In addition, we propose to observe the crystals using cold alkali atoms, by weakly dressing the groundstate of each atom with an excited Rydberg state with large dipole moment $d_r$, induced by a strong DC electric field $\mathcal{E_{\rm DC}}$ directed perpendicular to the 2D plane. For large interparticle separations $r \gg 100$ nm, outside of the dipole-blockaded region~\cite{RydbergExp,Santos}, the effective interatomic interaction is well described by the dipole-dipole potential with an effective dipole-moment $d\sim (\Omega/\Delta)^2 d_r$, where $\Omega$ and $\Delta$ are the laser Rabi frequency and detuning from resonance, respectively. The effective single-particle spontaneous emission rate is $\Gamma\sim(\Omega/\Delta)^2\Gamma_r$, with $\Gamma_r$ the bare value. If one translates the latter into an effective heating rate, we expect the crystalline phases ($\tau\gg1$) to be observable up to an average ``melting'' time $\mathcal{T}\sim  T_{\rm M}/(\Gamma_{\rm eff}E_{R})$, with $E_R$ the photon recoil energy (in the tens of kHz), and $T_{\rm M} \sim 0.1 \epsilon_0$ the estimated melting temperature in the trap \cite{Kalia81}. For example, the groundstate of $^{87}$Rb atoms can be weakly coupled to the Rydberg-state ($n=20$, $n_1=19$, $n_2=0$) with bare dipole moment $d_r\approx1.45{\rm kD}$ and $\Gamma_r/2 \pi \sim 100 $kHz.  Here, $n_1$ and $n_2$ are parabolic quantum numbers~\cite{GallagherBook}. For a DC field $\mathcal{E_{\rm DC}}=25$kV/m and a coupling laser with $\Omega/2 \pi=80{\rm MHz}$ and $\Delta/2\pi=1{\rm GHz}$, we obtain $\Gamma/2\pi \approx 640$Hz, $E_R\approx h25$kHz and $d\approx9.1{\rm Debye}$. For in-plane confinement $\omega/2\pi= 1{\rm kHz}$, this gives $r_0\approx 1.07 \mu{\rm m}$, $\tau\approx10$ and $\mathcal{T}\approx 62.5\mu{\rm s}$. To check the estimate for $\mathcal{T}$ we performed molecular dynamics simulations for the 2D system in the classical limit ($\tau\rightarrow\infty$), where the spontaneous emission was simulated by applying random in-plane kicks to each particle with momentum $(2mE_R)^{1/2}$ at an average rate $\Gamma$. We found the average energy $\langle E(t)\rangle$ to increase linearly with time $t$. For the parameters above and $N=13$ particles, the value $\langle E(t)-  E(0)\rangle/2N \approx 0.1\times\epsilon_0$ is reached at $t\approx 20\mu{\rm s}$ (roughly corresponding to a single kick) consistent with the above estimate for $\mathcal{T}$. However, the actual lifetime (corresponding to the thermal equilibration of the system) is found numerically to be of the order of ${\mathcal T} \simeq 200\mu$s, which is (much) larger than ${\mathcal T}$~\cite{MicheliUnpublished}.  

Since the in-situ interparticle distances can be of the order of a $\mu$m or more, it may be possible to directly address single particles {\it in situ}, and thus image the {\it spatial structure} of the crystalline phases above using, e.g., tightly focused beams. Alternatively, we propose the following method, which amounts to a cold-atom version of a {\it magnifying lens}. At a given time $t_0$ the (DC or AC) fields inducing the dipole-dipole interactions are switched off, and the in-plane harmonic confinement is inverted in sign. Because of this inverted potential, each particle experiences a radial acceleration which depends on its {\em spatial position} at time $t_0$. After a certain time-of-flight, the particles can be, e.g, ionized and their positions recorded on a ion plate with unit efficiency, providing a magnified picture of the in-situ spatial configuration.

The authors thank H.P.~B\"uchler and J.~Doyle for discussions. This
work was supported by IQOQI, the Austrian FWF, the EU through the STREP FP7-ICT-2007-C project NAME-QUAM, and the Canadian NSERC through the grant G121210893.


\begin{thebibliography}{40}
\bibitem{note} Obviously, any finite cluster can rotate without dissipation at a temperature lower than $\sim \hbar^2/I$, $I$ being the classical moment of inertia of the cluster.  We have observed the onset of such a regime for clusters with $\tau=5$, by noticing that the distribution of exchange cycles shows two sharp peaks in correspondence to the number of particles in the inner and outer rings, other cycles being absent. This is a conceptually different scenario than the one referred to here as ``superfluid".
\end{thebibliography}
\end{document}